\documentclass[12pt]{article}
\begin{document}
\title{From the Dark Matter Universe to the Dark Energy Universe}
\author{B.G. Sidharth\\
Dipartimento di Matematica e iInformatica,\\ Universita degli
Studi di Udine,\\ Via delle Scienza 206,Udine 33100 (Italy)}

\date{}
\maketitle
\begin{abstract}
Till the late nineties the accepted cosmological model was that of
a Universe that had originated in the Big Bang and was now
decelerating under the influence of as yet undetected dark matter,
so that it would come to a halt and eventually collapse. In 1997
however, the author had put forward a contra model wherein the
Universe was driven by dark energy, essentially the quantum zero
point field, and was accelerating with a small cosmological
constant. There were other deductions too, all in total agreement
with observation. All this got confirmation in 1998 and subsequent
observations have reconfirmed the findings.
\end{abstract}
\section{The New Cosmos}
When Einstein proposed his General Theory of Relativity early in
the last century, the accepted picture of the Universe was one
where all major constituents were stationary. This had puzzled
Einstein, because the gravitational pull of these constituents
should make the Universe collapse as the nett force would be
directed inwards. So he invented his famous cosmological constant,
essentially a repulsive force that would counterbalance the
attractive gravitational
force.\\
Shortly thereafter, two dramatic discoveries completely
transformed that picture. The first was due to astronomer Edwin
Hubble, who discovered that the basic constituents or building
blocks of the Universe were not stars, but rather, huge
conglomerations of stars, called galaxies. The second discovery,
aided by the redshift observations of the light of the galaxies
was the fact that these galaxies are rushing away from each other.
Rather than being static, the Universe is exploding. There was no
need for the counterbalancing cosmic repulsion any more and
Einstein dismissed
his proposal as his greatest blunder.\\
By the end of the last century, the Big Bang Model had been worked
out. It contained a huge amount of unobserved, hypothesized "matter"
of a new kind - dark matter. This was postulated as long back as the
1930s to explain the fact that the velocity curves of the stars in
the galaxies did not fall off, as they should. Instead they
flattened out, suggesting that the galaxies contained some
undetected and therefore non-luminous or dark matter. The identity
of this dark matter has been a matter of guess work, though. It
could consist of Weakly Interacting Massive Particles (WIMPS) or
Super Symmetric partners of existing particles. Or heavy neutrinos
or monopoles or unobserved brown dwarf stars and so on. In fact
Prof. Abdus Salam speculated some two decades ago \cite{salamnap}
"And now we come upon the question of dark matter which is one of
the open problems of cosmology. This is a problem which was
speculated upon by Zwicky fifty years ago. He showed that visible
matter of the mass of the galaxies in the Coma cluster was
inadequate to keep the galactic cluster bound. Oort claimed that the
mass necessary to keep our own galaxy together was at least three
times that concentrated into observable stars. And this in turn has
emerged as a central problem
of cosmology.\\
"You see there is the matter which we see in our galaxy. This is
what we suspect from the spiral character of the galaxy keeping it
together. And there is dark matter which is not seen at all by any
means whatsoever. Now the question is what does the dark matter
consist of? This is what we suspect should be there to keep the
galaxy bound. And so three times the mass of the matter here in our
galaxy should be around in the form of the invisible matter. This is
one of the
speculations."\\
The universe in this picture, contained enough of the mysterious
dark matter to halt the expansion and eventually trigger the next
collapse. It must be mentioned that the latest WMAP survey
\cite{science2}, in a model dependent result indicates that as much
as twenty three percent of the Universe is made up of dark matter,
though there is no definite observational confirmation
of its existence.\\
That is, the Universe would expand
up to a point and then collapse.\\
There still were several subtler problems to be addressed. One was
the famous horizon problem. To put it simply, the Big Bang was an
uncontrolled or random event and so, different parts of the Universe
in different directions were disconnected at the very earliest stage
and even today, light would not have had enough time to connect
them. So they need not be the same. Observation however shows that
the Universe is by and large uniform, rather like people in
different countries showing the same habits or dress. That would not
be possible without some form of faster than light
intercommunication which would violate Einstein's Special
Theory of Relativity.\\
The next problem was that according to Einstein, due to the material
content in the Universe, space should be curved whereas the Universe
appears to be flat. There were other problems as well. For example
astronomers predicted that there should be monopoles that is, simply
put, either only North magnetic poles or only South magnetic poles,
unlike the North South combined magnetic poles we encounter. Such
monopoles have
failed to show up even after seventy five years.\\
Some of these problems as we noted, were sought to be explained by
what has been called inflationary cosmology whereby, early on, just
after
the Big Bang the explosion was super fast \cite{zee,lindepl82}.\\
What would happen in this case is, that different parts of the
Universe, which could not be accessible by light, would now get
connected. At the same time, the super fast expansion in the initial
stages would smoothen out any distortion or curvature effects in
space,
leading to a flat Universe and in the process also eliminate the monopoles.\\
Nevertheless, inflation theory has its problems. It does not seem to
explain the cosmological constant observed since. Further, this
theory seems to imply that the fluctuations it produces should
continue to indefinite distances. Observation seems to imply the
contrary.\\
One other feature that has been studied in detail over the past few
decades is that of structure formation in the Universe. To put it
simply, why is the Universe not a uniform spread of matter and
radiation? On the contrary it is very lumpy with planets, stars,
galaxies and so on, with a lot of space separating these objects.
This has been explained in terms of fluctuations in density, that
is, accidentally more matter being present in a given region.
Gravitation would then draw in even more matter and so on. These
fluctuations would also cause the cosmic background radiation to be
non uniform or anisotropic. Such anisotropies are in fact being
observed. But this is not the end of the story. The galaxies seem to
be arranged along two dimensional structures and filaments with huge
separating
voids. \\
From 1997, the conventional wisdom of cosmology that had
concretized from the mid sixties onwards, began to be challenged.
It had been believed that the density of the Universe is near its
critical value, separating eternal expansion and ultimate
contraction, while the nuances of the dark matter theories were
being fine tuned. But that year, the author proposed a contra
view. To put it briefly, the universe is permeated by a
background dark energy, the Quantum Zero Point Field.\\
There would be fluctuations in this all permeating Zero Point Field
- or dark energy in the process of which, particles would be created
\cite{bgsfqp,bgsmg8,ijmpa,ijtp}. This model while consistent with
astrophysical observations predicted an ever expanding and
accelerating Universe with a small cosmological constant. It deduces
from theory the so called Large Number coincidences including the
purely empirical Weinberg formula that connects the pion mass to the
Hubble Constant \cite{narlikarcos,weinberggc} -- ''coincidences''
that have
troubled and mystified scientists from time to time.\\
However the work of Perlmutter and others \cite{perlnature,kirshner}
began appearing in 1998 and told a different story. These
observations of distant supernovae indicated that contrary to widely
held belief, the Universe was not only not decelerating, it was
actually accelerating though slowly. All this was greeted by the
community with skepticism $--$ either it was plain wrong,
or, let us wait and see.\\
A $2000$ article in the Scientific American \cite{musser} observed,
"In recent years the field of cosmology has gone through a radical
upheaval. New discoveries have challenged long held theories about
the evolution of the Universe... Now that observers have made a
strong case for cosmic acceleration, theorists must explain it....
If the recent turmoil is anything to go by, we had better keep
our options open."\\
On the other hand, an article in Physics World in the same year
noted \cite{caldwell}, "A revolution is taking place in cosmology.
New ideas are usurping traditional notions about the composition of
the Universe, the relationship between geometry
and destiny, and Einstein's greatest blunder."\\
The infamous cosmological constant was resurrected and now it was
"dark energy"
that was in the air, rather than dark matter. The universe had taken a U turn.\\
Let us now examine this new cosmology and some of its implications.
We will first go over the essentials and then examine the nuances.
\section{The Mysterious Dark Energy}
We first observe that the concept of a Zero Point Field (ZPF) or
Quantum vacuum (or  Aether) is an idea whose origin can be traced
back to Max Planck himself. Quantum Field Theory attributes the ZPF
to the virtual Quantum effects of an already present electromagnetic
field \cite{bd}. What is the mysterious energy of supposedly
empty vacuum? \cite{duffy}.\\
It may sound contradictory to attribute energy or density to the
vacuum. After all vacuum in the older concept is a total void.
However, over the past four hundred years, it has been realized that
it may be necessary to replace the vacuum by a medium with some
specific physical properties. These properties were chosen to suit
the specific requirements of the time. For instance Descartes the
seventeenth century French philosopher mathematician proclaimed that
the so called empty space above the mercury column in a Torricelli
tube, that is, what is called the Torricelli vacuum, is not a vacuum
at all. Rather, he said,
it was something which was neither mercury nor air, something he called  aether.\\
The seventeenth century Dutch Physicist, Christian Huygens required
such a non intrusive medium like  aether, so that light waves could
propagate through it, rather like the ripples on the surface of a
pond. This was the luminiferous  aether. In the nineteenth century
the aether was reinvoked. Firstly in a very intuitive way Faraday
could conceive of magnetic effects in vacuum in connection with his
experiments on induction. Based on this, the  aether was used for
the propagation of electromagnetic waves in Maxwell's Theory of
electromagnetism, which in fact laid the stage for Special
Relativity. This  aether was a homogenous, invariable,
non-intrusive, material medium which could be used as an absolute
frame of reference at least for certain chosen observers. The
experiments of Michelson and Morley towards the end of the
nineteenth century were sought to be explained in terms of aether
that was dragged by bodies moving in it. Such explanations were
untenable and eventually lead to its downfall, and thus was born
Einstein's Special Theory of Relativity in which there is no such
absolute
frame of reference. The  aether lay discarded once again.\\
Very shortly thereafter the advent of Quantum Mechanics lead to its
rebirth in a new and unexpected avatar. Essentially there were two
new ingredients in what is today called the Quantum vacuum. The
first was a realization that Classical Physics had allowed an
assumption to slip in unnoticed: In a source or charge free
"vacuum", one solution of Maxwell's Equations of electromagnetic
radiation is no doubt the zero solution. But there is also a more
realistic non zero solution. That is, the
electromagnetic radiation does not necessarily vanish in empty space.\\
The second ingredient was the mysterious prescription of Quantum
Mechanics, the Heisenberg Uncertainty Principle, according to which
it would be impossible to precisely assign momentum and energy on
the one hand and spacetime location on the other. Clearly the
location of a vacuum with no energy or momentum cannot be specified in spacetime.\\
This leads to what is called a Zero Point Field. For instance a
Harmonic oscillator, a swinging pendulum for example, according to
classical ideas has zero energy and momentum in its lowest
position. But the Heisenberg Uncertainty endows it with a
fluctuating energy. This fact was recognized by Einstein himself
way back in 1913, who contrary to popular belief, retained the
concept of  aether though from a different perspective
\cite{wilczek}. It also provides an understanding of the
fluctuating electromagnetic field in vacuum.
that this can be modeled by a Weiner process.\\
From another point of view, according to classical ideas, at the
absolute zero of temperature, there should not be any motion. After
all the zero is when all thermodynamic motion ceases. But as Nernst,
father of the third law of Thermodynamics himself noted,
experimentally this is not so. There is the well known superfluidity
due to Quantum Mechanical -- and not thermodynamic -- effects. This
is the situation where supercooled
Helium moves in a spooky fashion.\\
This mysterious Zero Point Field or Quantum vacuum energy has since
been experimentally confirmed in effects like the  Casimir effect
which demonstrates a force between uncharged parallel plates
separated by a charge free medium, the Lamb shift which demonstrates
a minute jiggling of an electron orbiting the nucleus in an atom
$--$ as if it was being buffeted by the Zero Point Field, and as we
will see, the anomalous Quantum Mechanical gyromagnetic ratio $g =
2$, the Quantum Mechanical spin half and so
on \cite{milonniqv}-\cite{podolny}, \cite{mwt}.\\
The Quantum vacuum is a far cry however, from the passive  aether of
olden days. It is a violent medium in which charged particles like
electrons and positrons are constantly being created and destroyed,
almost instantly, in fact within the limits permitted by the
Heisenberg Uncertainty Principle for the violation of energy
conservation. One might call the Quantum vacuum as a new state of
matter, a compromise between something and nothingness. Something
which corresponds to what the Rig Veda described thousands
of years ago: "Neither existence, nor non existence."\\
Quantum vacuum can be considered to be the lowest state of any
Quantum field, having zero momentum and zero energy. The fluctuating
energy or ZPF due to Heisenberg's principle has an infinite value
and is "renormalized", that is, discarded. The properties of the
Quantum vacuum can under certain conditions be altered, which was
not the case with the erstwhile aether. In modern Particle Physics,
the Quantum vacuum is responsible for apart from effects alluded to
earlier, other phenomena like quark confinement, a property we
already encountered, whereby it would be impossible to observe an
independent or free quark, the spontaneous breaking of symmetry of
the electro weak theory, vacuum polarization wherein charges like
electrons are surrounded by a cloud of other opposite charges
tending to mask the main charge and so on. There could be regions of
vacuum fluctuations comparable to the domain structures of
ferromagnets. In a ferromagnet, all elementary electron-magnets are
aligned with their spins in a certain direction.
However there could be special regions wherein the spins are aligned differently.\\
Such a Quantum vacuum can be a source of cosmic repulsion, as
pointed by Zeldovich and others \cite{zeldovich,cu}. However a
difficulty in this approach has been that the value of the
cosmological constant turns out to be huge, far beyond what is
observed. This has been
called the cosmological constant problem \cite{weinbergprl}.\\
There is another approach, Stochastic Electrodynamics which treats
the ZPF as independent and primary and attributes to it Quantum
Mechanical effects \cite{santos,depenastoch}. It may be
re-emphasized that the ZPF results in the well known
experimentally verified Casimir effect
\cite{mostepanenko,lamoreauz}. We would also like to point out
that contrary to popular belief, the concept of aether has
survived over the decades through the works of Dirac, Vigier,
Prigogine, String Theorists like Wilzeck and others
\cite{hushwater},\cite{petronivig}-\cite{leequarks}. As pointed
out it appears that even Einstein himself
continued to believe in this concept \cite{achuthan}.\\
We would first like to observe that the energy of the fluctuations
in the background electromagnetic field could lead to the formation
of elementary particles. Indeed this was Einstein's belief. As
Wilzeck (loc.cit) put it, ``Einstein was not satisfied with the
dualism. He wanted to regard the fields, or ethers, as primary. In
his later work, he tried to find a unified field theory, in which
electrons (and of course protons, and all other particles) would
emerge as solutions in which energy was especially concentrated,
perhaps as singularities. But his efforts in this direction did not
lead to
any tangible success.''\\
We will now argue that indeed this can happen. In the words of
Wheeler \cite{mwt}, ``From the zero-point
 fluctuations of a single oscillator to the fluctuations of the electromagnetic field to
 geometrodynamic fluctuations is a natural order of progression...''\\
Let us consider, following Wheeler a harmonic oscillator in its
ground state. The probability amplitude is
$$\psi (x) = \left(\frac{m\omega}{\pi \hbar}\right)^{1/4} e^{-(m\omega/2\hbar)x^2}$$
for displacement by the distance $x$ from its position of classical
equilibrium. So the oscillator fluctuates over an interval
$$\Delta x \sim (\hbar/m\omega)^{1/2}$$
The electromagnetic field is an infinite collection of independent
oscillators, with amplitudes $X_1,X_2$ etc. The probability for the
various oscillators to have amplitudes $X_1, X_2$ and so on is the
product of individual oscillator amplitudes:
$$\psi (X_1,X_2,\cdots ) = exp [-(X^2_1 + X^2_2 + \cdots)]$$
wherein there would be a suitable normalization factor. This
expression gives the probability amplitude $\psi$ for a
configuration $B (x,y,z)$ of the magnetic field that is described by
the Fourier coefficients $X_1,X_2,\cdots$ or directly in terms of
the magnetic field configuration itself by
$$\psi (B(x,y,z)) = P exp \left(-\int \int \frac{\bf{B}(x_1)\cdot \bf{B}(x_2)}{16\pi^3\hbar cr^2_{12}} d^3x_1 d^3x_2\right).$$
$P$ being a normalization factor. Let us consider a configuration
where the magnetic field is everywhere zero except in a region of
dimension $l$, where it is of the order of $\sim \Delta B$. The
probability amplitude for this configuration would be proportional
to
$$\exp \left[-\left((\Delta B)^2 l^4/\hbar c\right)\right]$$
So the energy of fluctuation in a volume of length $l$ is given by
finally \cite{mwt,bgsfqv,bgscosfluc}
\begin{equation}
B^2 \sim \frac{\hbar c}{l}\label{3eA}
\end{equation}
We will return to (\ref{3eA}) subsequently but observe that if in
(\ref{3eA}) above $l$ is taken to be the Compton wavelength of a
typical elementary particle, then we recover its energy $mc^2$, as
can be easily verified. We had previously seen how inertial mass
and energy can be deduced on the basis of viscous resistance to
the ZPF.  We will also deduce this from Quantum Mechanical effects
within the Compton scale. The above gives us back this result in
the context of the ZPF. In any case (\ref{3eA})
shows the inverse dependence of the length scale and the energy (or momentum).\\
It may be reiterated that Einstein himself had believed that the
electron was a result of such a condensation from the background
electromagnetic field (Cf.\cite{castell,cu} for details). The above
result is very much in this spirit. In the sequel we also take the
pion to represent a typical elementary
particle, as in the literature.\\
To proceed, as there are $N \sim 10^{80}$ such particles in the
Universe, we get, consistently,
\begin{equation}
Nm = M\label{3e1}
\end{equation}
where $M$ is the mass of the Universe. It must be remembered that
the energy of gravitational interaction between the particles is
very much insignificant compared to the above electromagnetic considerations.\\
In the following we will use $N$ as the sole cosmological parameter.\\
We next invoke the well known relation
\cite{bgsfluc,nottalefractal,hayakawa}
\begin{equation}
R \approx \frac{GM}{c^2}\label{3e2}
\end{equation}
where $M$ can be obtained from (\ref{3e1}). We can arrive at
(\ref{3e2}) in different ways. For example, in a uniformly expanding
Friedman Universe, we have
$$\dot{R}^{2} = 8 \pi G\rho R^2/3$$
In the above if we substitute $\dot{R} = c$ at $R$, the radius of
the universe, we
get (\ref{3e2}). Another proof will be given later in Section 3.10.\\
We now use the fact that given $N$ particles, the
(Gaussian)fluctuation in the particle number is of the order
$\sqrt{N}$\cite{hayakawa,huang,ijmpa,ijtp,bgsfqp,bgsmg8}, while a
typical time interval for the fluctuations is $\sim \hbar/mc^2$,
the Compton time, the fuzzy interval of recent approaches,within
which there is no meaningful physics.  So particles are created
and destroyed - but the ultimate result is that $\sqrt{N}$
particles are created just as this is the nett displacement in a
random walk of unit step. So we have,
\begin{equation}
\frac{dN}{dt} = \frac{\sqrt{N}}{\tau}\label{3ex}
\end{equation}
whence on integration we get, (remembering that we are almost in the
continuum region that is, $\tau \sim 10^{-23}sec \approx 0$),
\begin{equation}
T = \frac{\hbar}{mc^2} \sqrt{N}\label{3e3}
\end{equation}
We can easily verify that the equation (\ref{3e3}) is indeed
satisfied where $T$ is the age of the Universe. Next by
differentiating (\ref{3e2}) with respect to $t$ we get
\begin{equation}
\frac{dR}{dt} \approx HR\label{3e4}
\end{equation}
where $H$ in (\ref{3e4}) can be identified with the Hubble Constant,
and using (\ref{3e2}) is given by,
\begin{equation}
H = \frac{Gm^3c}{\hbar^2}\label{3e5}
\end{equation}
Equation (\ref{3e1}), (\ref{3e2}) and (\ref{3e3}) show that in this
formulation, the correct mass, radius, Hubble constant and age of
the Universe can be deduced given $N$, the number of particles, as
the sole cosmological or large scale parameter. We observe that at
this stage we are not invoking any particular dynamics - the
expansion is due to the random creation of particles from the ZPF
background. Equation (\ref{3e5}) can be written as
\begin{equation}
m \approx \left(\frac{H\hbar^2}{Gc}\right)^{\frac{1}{3}}\label{3e6}
\end{equation}
Equation (\ref{3e6}) has been empirically known as an "accidental"
or "mysterious" relation. As observed by Weinberg \cite{weinberggc},
this is unexplained: it relates a single cosmological parameter $H$
to constants from microphysics. We will touch upon this micro-macro
nexus again. In our formulation, equation (\ref{3e6}) is no longer a
mysterious coincidence but
rather a consequence of the theory.\\
As (\ref{3e5}) and (\ref{3e4}) are not exact equations but rather,
order of magnitude relations, it follows, on differentiating
(\ref{3e4}) that a small cosmological constant $\wedge$ is allowed
such that
$$\wedge \leq 0 (H^2)$$
This is consistent with observation and shows that $\wedge$ is very
small $--$ this has been a puzzle, the so called cosmological
constant problem alluded to, because in conventional theory, it
turns out to be huge \cite{weinbergprl}. But it poses no problem in
this formulation. This is because of the characterization of the ZPF
as independent and primary in our formulation this being the
mysterious dark energy. \\
To proceed we observe that because of the fluctuation of $\sim
\sqrt{N}$ (due to the ZPF), there is an excess electrical potential
energy of the electron, which in fact we identify as its inertial
energy. That is \cite{ijmpa,hayakawa},
$$\sqrt{N} e^2/R \approx mc^2.$$
On using (\ref{3e2}) in the above, we recover the well known
Gravitation-Electromagnetism ratio viz.,
\begin{equation}
e^2/Gm^2 \sim \sqrt{N} \approx 10^{40}\label{3e7}
\end{equation}
or without using (\ref{3e2}), we get, instead, the well known so
called Weyl-Eddington formula,
\begin{equation}
R = \sqrt{N}l\label{3e8}
\end{equation}
(It appears that (\ref{3e8}) was first noticed by H. Weyl
\cite{singh}). Infact (\ref{3e8}) is the spatial counterpart of
(\ref{3e3}). If we combine (\ref{3e8}) and (\ref{3e2}), we get,
\begin{equation}
\frac{Gm}{lc^2} = \frac{1}{\sqrt{N}} \propto T^{-1}\label{3e9}
\end{equation}
where in (\ref{3e9}), we have used (\ref{3e3}). Following Dirac
(cf.also \cite{melnikov}) we treat $G$ as the variable, rather than
the quantities $m, l, c \,\mbox{and}\, \hbar$ which we will call
micro physical constants because of their central role
in atomic (and sub atomic) physics.\\
Next if we use $G$ from (\ref{3e9}) in (\ref{3e5}), we can see that
\begin{equation}
H = \frac{c}{l} \quad \frac{1}{\sqrt{N}}\label{3e10}
\end{equation}
Thus apart from the fact that $H$ has the same inverse time
dependance on $T$ as $G$, (\ref{3e10}) shows that given the
microphysical constants, and
$N$, we can deduce the Hubble Constant also, as from (\ref{3e10}) or (\ref{3e5}).\\
Using (\ref{3e1}) and (\ref{3e2}), we can now deduce that
\begin{equation}
\rho \approx \frac{m}{l^3} \quad \frac{1}{\sqrt{N}}\label{3e11}
\end{equation}
Next (\ref{3e8}) and (\ref{3e3}) give,
\begin{equation}
R = cT\label{3e12}
\end{equation}
Equations (\ref{3e11}) and (\ref{3e12}) are consistent with observation.\\
Finally, we observe that using $M,G \mbox{and} H$ from the above, we
get
$$M = \frac{c^3}{GH}$$
This relation is required in the Friedman model of the expanding
Universe (and the Steady State model too). In fact if we use in this
relation, the expression,
$$H = c/R$$
which follows from (\ref{3e10}) and (\ref{3e8}), then we recover
(\ref{3e2}). We will be repeatedly using these relations in the sequel.\\
 As we saw the above model predicts a dark
energy driven ever expanding and accelerating Universe with a small
cosmological constant while the density keeps decreasing. Moreover
mysterious large number relations like (\ref{3e5}), (\ref{3e11}) or
(\ref{3e8}) which were considered to be miraculous accidents now
follow from the underlying theory. This seemed to go against the
accepted idea that the density of the Universe equalled the critical
density required for closure and that aided by dark matter, the
Universe was decelerating. However, as noted, from 1998 onwards,
following the work of Perlmutter, Schmidt
and co-workers, these otherwise apparently heretic conclusions have been vindicated.\\
It may be mentioned that the observational evidence for an
accelerating Universe was the American Association for Advancement
of Science's Breakthrough of the Year, 1998 while the evidence for
nearly seventy five percent of the Universe being Dark Energy, based
on the Wilkinson Microwave Anisotropy Probe (WMAP) and the Sloan Sky
Digital Survey was the Breakthrough of the Year, 2003
\cite{science1,science2}.\\
In this case, particles are like the Benard cells which form in
fluids, as a result of a phase transition. While some of the
particles or cells may revert to the Zero Point Field, on the whole
there is a creation of $\sqrt{N}$ of these particles. If the average
time for the creation of the $\sqrt{N}$ particles or cells is
$\tau$, then at any point of time where there are $N$ such
particles, the time elapsed, in our case the age of the Universe,
would be given by (\ref{3e3}). While this is not exactly the Big
Bang scenario, there is nevertheless a rapid creation of matter from
the background Quantum vacuum or Zero Point Field. Thus over
$10^{40}$ particles would have been
created within a fraction of a second.\\
In any case when $\tau \to 0$, we recover the Big Bang scenario with
a singular creation of matter, while when $\tau \to$ Planck time we
recover the Prigogine Cosmology (Cf.\cite{cu} for details). However
in neither of these two limits we can deduce all the above
consistent with observation Large Number relations with the
cosmological constant $\Lambda$, which therefore have then to be
branded as accidents.\\
The above scheme which throws up a time varying gravitational
constant as in the case of Dirac or Brans-Dicke cosmologies explains
all the standard results like the precession of the perihelion of
Mercury and so on and also explains the anomalous accelerations of
the Pioneer spacecrafts and shortening of the orbital period of
binary pulsars and other recent effects. In this scheme there is no
need to invoke dark matter, which plays a marginal role
(Cf.\cite{tduniv} for details).\\
Finally, it may be pointed out that recently Krauss has observed
that the cosmic acceleration means that the details of the
beginnings of the Universe will be lost in future. As we may not be
living at a privileged point of time, the question may be asked,
have we already lost these details.

\end{document}